\documentclass[%
 reprint,superscriptaddress,
 amsmath,amssymb,
 aps,
 pra,
]{revtex4-2}
\usepackage{graphicx}
\usepackage{dcolumn}
\usepackage{bm}

\usepackage[colorlinks=true, allcolors=blue]{hyperref}

\begin{document}

\preprint{APS/123-QED}

\title{A large magneto-optical trap of cadmium atoms loaded from a cryogenic buffer gas beam}


\author{J. E. Padilla-Castillo}
\thanks{Authors contributed equally}
\affiliation{Fritz-Haber-Institut der Max-Planck-Gesellschaft, Faradayweg 4-6, 14195 Berlin, Germany}
\author{S. Hofs\"{a}ss}
\thanks{Authors contributed equally}
\affiliation{Fritz-Haber-Institut der Max-Planck-Gesellschaft, Faradayweg 4-6, 14195 Berlin, Germany}
\author{L. Pal\'{a}nki}
\affiliation{Centre for Cold Matter, Blackett Laboratory, Imperial College London, London SW7 2AZ}
\author{J. Cai}
\affiliation{Fritz-Haber-Institut der Max-Planck-Gesellschaft, Faradayweg 4-6, 14195 Berlin, Germany}
\author{C. J. H. Rich}
\affiliation{Centre for Cold Matter, Blackett Laboratory, Imperial College London, London SW7 2AZ}
\author{R. Thomas}
\affiliation{Fritz-Haber-Institut der Max-Planck-Gesellschaft, Faradayweg 4-6, 14195 Berlin, Germany}
\author{S. Kray}
\affiliation{Fritz-Haber-Institut der Max-Planck-Gesellschaft, Faradayweg 4-6, 14195 Berlin, Germany}
\author{G. Meijer}
\affiliation{Fritz-Haber-Institut der Max-Planck-Gesellschaft, Faradayweg 4-6, 14195 Berlin, Germany}
\author{S. C. Wright}
\affiliation{Fritz-Haber-Institut der Max-Planck-Gesellschaft, Faradayweg 4-6, 14195 Berlin, Germany}
\author{S. Truppe}
\email[]{s.truppe@imperial.ac.uk}
\affiliation{Fritz-Haber-Institut der Max-Planck-Gesellschaft, Faradayweg 4-6, 14195 Berlin, Germany}
\affiliation{Centre for Cold Matter, Blackett Laboratory, Imperial College London, London SW7 2AZ}

\date{\today}

\begin{abstract}
We demonstrate rapid loading of a magneto-optical trap (MOT) of cadmium atoms from a pulsed cryogenic helium buffer gas beam, overcoming strong photoionization losses. Using the $ ^1S_0 \rightarrow{} ^1P_1 $ transition at 229 nm, we capture up to $ 1.1(2) \times 10^7$ $^{112}$Cd atoms in 10 ms, achieving a peak density of $2.5 \times 10^{11}$cm$^{-3}$ and a phase-space density of $ 2 \times 10^{-9} $. The large scattering force in the deep ultraviolet enables Zeeman slowing within 5 cm of the trap, yielding a capture velocity exceeding 200 m/s. We measure the MOT trap frequency and damping constant, and determine the absolute photoionization cross section of the $^1P_1 $ state. Photoionization losses are mitigated via dynamic detuning of the trapping light's frequency, allowing efficient accumulation of multiple atomic pulses. Our results demonstrate the benefits of deep-UV (DUV) transitions and cryogenic beams for loading high-density MOTs, especially for species with significant loss channels in their main cooling cycle. The cadmium MOT provides a robust testbed that benchmarks our DUV laser cooling system and establishes the foundation for trapping and cooling polar AlF molecules, which share many optical and structural properties with Cd.

\end{abstract}

\maketitle


\section{\label{sec:level1}Introduction}


Laser-cooled alkaline-earth and alkaline-earth-like atoms are widely used for applications in optical atomic clocks \cite{Ludlow2015, boulder_atomic_clock_optical_network_bacon_collaboration_frequency_2021, mcferran_erratum_2015, ohmae_direct_2020}, atom interferometry \cite{Ruschewitz1998, Poli2011, abe_matter-wave_2021, tino_testing_2021, bandarupally_design_2023}, quantum information simulation and computation \cite{Daley2008, gorshkov_two-orbital_2010, daley_quantum_2011, pagano_fast_2019, madjarov_high-fidelity_2020, bloch_quantum_2012, schafer_tools_2020}, and the search for new physics beyond the Standard Model of particle physics \cite{Berengut2018, Safranova2018}. Interest in laser cooling Zn \cite{Buki21}, Cd \cite{Brickman2007, Kaneda2016, Schussheim18, Katori2019, Tinsley2021, Tinsley2022, Bandarupally2023, gibble_laser-cooling_2024} and Hg \cite{Lavigne2022} has grown due to their low sensitivity to black body radiation-induced ac Stark shifts \cite{dzuba_blackbody_2019}, a key source of uncertainty in optical lattice clocks \cite{Ludlow2015}. Clocks based on these atomic species promise to be accurate, compact, and portable \cite{Katori2019}. Cd, in particular, has six stable bosonic and two fermionic isotopes, making it a prime candidate for precise isotope shift spectroscopy to search for new physics beyond the Standard Model \cite{Ohayon2022}.

The $(5s^2)^1S_0\rightarrow\,(5s5p)^1P_1 $ (229~nm) transition in cadmium is the shortest transition wavelength magneto-optical trap (MOT) \cite{Brickman2007, Kaneda2016, Katori2019} to date, inspiring and guiding efforts to trap using analogous transitions in Zn (213.8~nm) \cite{Buki21}, Hg (185~nm) \cite{Lavigne2022} and the diatomic polar molecule AlF (227.5 nm) \cite{PhysRevA.100.052513, hofsass2021optical}. The combination of a short excited-state lifetime ($\tau=1.60(5)$ ns for the $^1P_1$ state of Cd \cite{Hofsaess2023}) with a short transition wavelength results in a large radiation pressure force. The minimum stopping distance for particles moving with an initial speed of 200 m/s is about 4 mm, enabling compact setups and the rapid loading of large MOTs. Moreover, the small photon-absorption cross-section in the deep ultraviolet (DUV) reduces radiation-trapping effects enabling high-density MOTs \cite{Katori2019}.

We choose to study Cd not only for its metrological interest, but also as a proxy for the laser-coolable molecule AlF, which shares similar DUV transitions while posing additional challenges due to a complex internal level structure \cite{hofsass2021optical}. Cd enables systematic studies of MOT loading, radiation pressure slowing, and accumulation dynamics that can inform future molecule-based experiments. Figure \ref{fig:figure1}a) illustrates the relevant energy-level structure of the Cd isotopes. The six bosonic isotopes are free of hyperfine structure, whereas the two fermionic isotopes, with nuclear spin $I_N = 1/2$, have a hyperfine splitting of approximately $290$ MHz in the $^1P_1$ excited state. The boson resonance lines are separated by $300-500$~MHz and partially overlap with those of the fermions; we recently reported accurate isotope shift and hyperfine structure measurements \cite{Hofsaess2023} and draw on these in our analysis of the MOT.

While conventional DUV Cd MOTs are loaded from a thermal vapor \cite{Brickman2007, Katori2019}, this loading method is fundamentally limited by photoionization losses from the $^1P_1$ state. One solution is to avoid excitation on the 229 nm transition entirely and instead trap atoms using a combination of the narrow 326 nm $^1S_0 \rightarrow$\,$ ^3P_1$ intercombination line with the broader 361 nm $^3P_2 \rightarrow$\,$^3D_3$ transition \cite{gibble_laser-cooling_2024}. A similar two-stage strategy has also been proposed and simulated in detail \cite{bandarupally_design_2023}. These approaches eliminate or minimize photoionization losses while retaining efficient laser cooling.

Here, we demonstrate an alternative solution: rapid and efficient loading of a Cd MOT from a cryogenic helium buffer gas-cooled, pulsed atomic beam. Buffer gas sources are rarely used to load atomic MOTs, but in certain circumstances can offer valuable advantages. One example is the loading of refractory elements, for which very high temperatures are required to generate a thermal beam; examples include Yb \cite{kuwamoto1999, Hemmerling2014}, Er \cite{McClelland2006, Hemmerling2014}, Cr \cite{Bradley2000}, Ho \cite{Miao2014, Hemmerling2014}, Tm \cite{Sukachev2010, Hemmerling2014} and Dy \cite{Lu2010} with many more transition metal species currently being considered \cite{eustice2020laser}. Another instance occurs when a loss channel in the laser cooling scheme limits the lifetime in a MOT, meaning that loading with a short, intense, pulse is highly desirable to maximize the number of trapped particles. This is especially beneficial when loading molecular MOTs, where losses to excited vibrational levels in the electronic ground state are unavoidable \cite{barry2014magneto,truppe2017molecules, anderegg2017radio,collopy2018,PhysRevLett.134.083401,vilas2022magneto,PhysRevLett.133.143404}. A further advantage of fast loading is the significant reduction in the experimental cycle time, by minimizing dead time and increasing the achievable repetition rate. 

Photoionization limits the achievable density and lifetime of cadmium MOTs operating in the DUV, particularly when loaded from a thermal vapor source \cite{Brickman2007}. In contrast, using an intense, pulsed, and slow buffer gas beam of Cd, we achieve a phase-space density of $2 \times 10^{-9}$ within just 10 ms, corresponding to a MOT loading rate exceeding $10^9\ \text{s}^{-1}$, 100 times faster than previously reported Cd MOTs \cite{Katori2019, gibble_laser-cooling_2024}. To mitigate photoionization losses, we detune the trapping light further from resonance immediately after loading, enabling the accumulation of multiple atomic pulses. The large radiation pressure when exciting the $^1S_0\rightarrow\,^1P_1$ also allows the MOT magnetic field to serve as an integrated Zeeman slower. Together, these features yield a highly compact system, with the combined trap and slower occupying just 0.2 liters.


\section{Experimental setup and simulations}
\begin{figure*}[]
    \centering
    \includegraphics[width=2 \columnwidth]{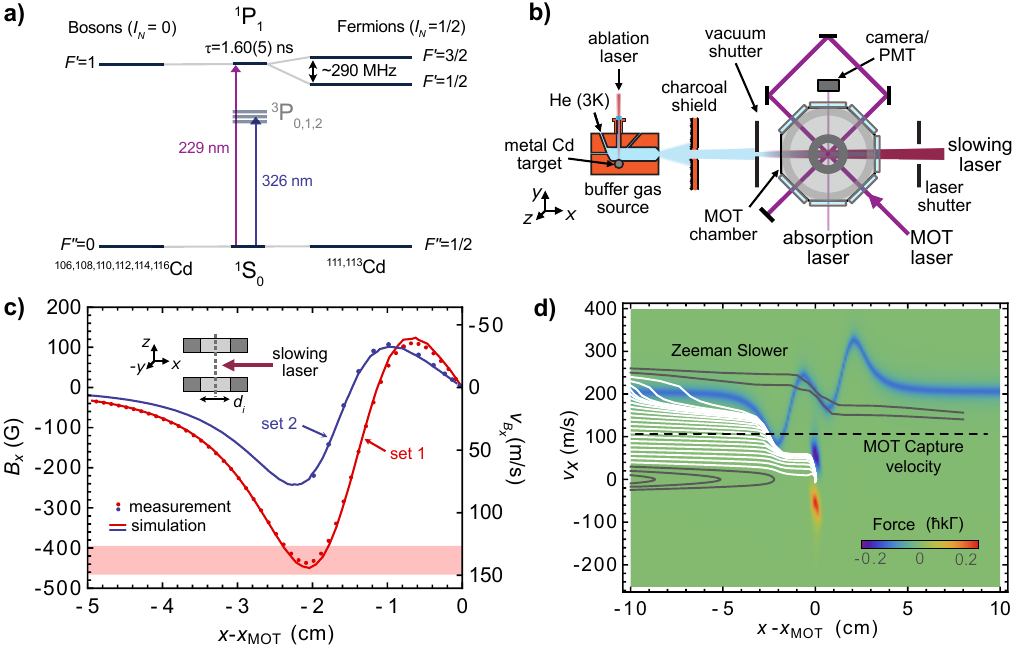}
    \caption{a) Diagram of the relevant Cd energy levels, for bosonic (left) and fermionic (right) isotopes. b) Experimental setup: Cd atoms are captured in a MOT from a laser-slowed, pulsed cryogenic helium buffer gas beam or from a background vapor. Fluorescence from the atomic beam and the MOT is imaged onto a photomultiplier tube or camera. An optional low-intensity probe beam is used to image absorption by the trapped atoms onto the camera. The magnetic field is produced by permanent ring magnets. c) Magnetic field profile of the trapping magnets showing the radial component $B_x$ along the $x$-axis near the MOT center. Measurements (points) are compared with a finite element simulation (solid curves). The right axis plots $v_{B_x} = \mu_B B_x/(h\lambda)$, which is the atomic velocity whose Doppler shift compensates the Zeeman shift of the state $m_{F'} = +1$, with the quantization axis along $x$. The red shaded bar indicates the span of the transition linewidth in these units. d) Simulations: pyLCP trajectory simulations of the Zeeman slower and MOT, with magnet set I. White trajectories indicate successful capture.}
    \label{fig:figure1}
\end{figure*}


As part of a close collaboration, we built two nearly identical experimental setups, the first at the Fritz Haber Institute (FHI) in Berlin, and a second at Imperial College London (ICL). Most experiments presented here have been performed at the FHI, while a subset have been reproduced at ICL. Figure \ref{fig:figure1}b) shows a schematic of the setup. We produce a pulsed Cd beam in a cryogenic helium buffer gas source \cite{Hutzler2012, Truppe2018} and load atoms into the MOT in a second, differentially pumped chamber. To characterize the MOT, we detect the fluorescence with either a DUV-sensitive photomultiplier tube (PMT), EM-CCD camera or a CMOS camera with a rolling shutter. The EM-CCD camera can also be used to image absorption of a low-intensity probe laser interacting with the trapped atoms. Without helium buffer gas flowing into the source, the pressure of the trap chamber reaches $1 \times 10^{-9}$~mbar. After MOT loading, an in-vacuum shutter blocks the helium, maintaining a steady-state pressure below $5 \times 10^{-9}$~mbar. If needed, the base pressure can be further reduced by improving the differential pumping between the source and trap chambers.

Each experiment begins with a pulse of ablation light from an Nd:YAG laser (1064 nm), which is gently focused onto a Cd metal target inside the buffer gas cell. The target comprises an Al holder that contains a Cd sample of natural isotopic abundance. Ablated Cd atoms are cooled with 3~K He buffer gas, and exit the cell as a short pulse ($\approx 1$~ms) of atoms with a forward velocity distribution centered around $130$~m/s and with a full width at half maximum of 60~m/s. Firing the ablation laser defines $t=0$.

Our MOT is placed 40~cm from the exit of the buffer gas cell, so that the typical flight time between the source and the trap is 3~ms. The required quadrupole magnetic field is generated by two permanent ferrite ring magnets with magnetization $\pm M\vec{z}$, whose centers are positioned at $\pm 12.5$~mm along the $z$-axis. We use magnets of thickness $9$~mm, outer diameter $40$~mm, and $|M|=290$~kA/m, and cut the inner diameter of the bore $d_i$ to adjust the field profile. Figure \ref{fig:figure1}c) shows the measured field component $B_x$ along the $x$-axis for the two different magnet sets compared to a finite element model. Magnet set I has $d_i = 22$~mm and a radial magnetic field gradient $A=dB_x/dx=250$~G/cm, shown in red in the figure; magnet set II has $d_i=30$~mm, resulting in $A=145$~G/cm. A dispenser source containing Cd with natural abundance is mounted about 10 cm from the center of the MOT chamber. This allows us to compare our buffer-gas-beam-loaded MOT with the more commonly used dispenser-loaded MOT \cite{Katori2019}. 


At FHI, we generate the laser light at 229~nm using two frequency-quadrupled continuous-wave titanium-sapphire (Ti:Sa) systems, each capable of generating up to 250~mW, of which we typically use $40-100$ mW. The fundamental light of each Ti:Sa near 915~nm is frequency stabilized to below 1~MHz via a HighFinesse WS8-10 wavemeter referenced to a calibrated, frequency-stabilized HeNe laser at 633~nm. The output mode of the 229~nm light is elliptical with an aspect ratio of about 3. 
At ICL, we frequency-quadruple vertical external cavity surface emitting lasers (Vexlum Valo SF) in custom-made doubling cavities (Agile Optic) to generate 250~mW near 229 nm. The fundamental light is stabilized to a WS8-10 wavemeter referenced to a laser whose frequency is locked to a transition in Rb with a wavelength close to 780~nm.

The MOT laser beam is shaped to be approximately Gaussian (TEM$_{00}$ mode) with a 1/$e^2$ diameter of 3.5~mm, and split into two paths at the experiment. One path provides the trapping light along $z$, and the other path provides the trapping light in the $x,y$ plane. This light circulates the chamber in a ``folded-beam" configuration, before being retro-reflected (see Figure \ref{fig:figure1}b). This beam configuration is a compromise between maximizing the total intensity of the trapping light, and sufficient balancing of the forward and retro-reflected laser intensities. By minimizing the number of optical elements between the laser output and the MOT, we reduce unwanted birefringence, losses, and beam distortions.

We estimate the total peak intensity at the centre of the trap, including transmission losses, to be $I=2.2$~W/cm$^2$. Optical shutters outside the vacuum chamber block the laser light when not needed, to reduce DUV-induced damage to the vacuum viewports.

The second laser provides the slowing light, which counter-propagates relative to the atomic beam. We use approximately 40~mW of laser power and a peak intensity of $0.4$~W$/$cm$^2$. This light is turned off at $t=10$~ms after loading the MOT. An important aspect of the setup is the effect of the slowing laser in the region 5~cm from the MOT center, in the exterior region of the ring magnets, as shown in figure \ref{fig:figure1}c). The scale of the right axis shows $v_{B_x} = \mu_B B_x m_{F'} \lambda/h$, which is the velocity along $x$ required to compensate the Zeeman shift of the $^1P_1$ excited states and ensure that a Cd atom remains in resonance with a laser traveling along $x$. We see that in the region $-5$~cm$\,<(x-x_{\mathrm{MOT}})<-2$~cm, the Zeeman shift of magnet set I is sufficient to cover a velocity range of 140~m/s, and enough to bring the fastest atoms from the source below 50~m/s. The shaded red bar in the figure has a width $\Gamma/k = 23$~m/s and  represents the velocity range within $\pm\Gamma/2$ of the resonance condition. Here, $\Gamma=1/\tau$ is the spontaneous decay rate and $k=2\pi/\lambda$, with $\lambda=228.87$ nm. The broad transition linewidth ($\Gamma/(2\pi)=99.7(3.4)$ MHz \cite{Hofsaess2023}) significantly relaxes constraints on the magnetic field profile for effective Zeeman slowing. 

To help understand the performance of our MOT, we implemented trajectory simulations using the pyLCP code package \cite{Eckel2022}. The simulations contain isotope-specific information such as the transition isotope shifts, hyperfine structure and relative natural abundances, the three-dimensional profile of the trapping magnets, and the laser geometry as illustrated in Figures \ref{fig:figure1}b) and c). Figure \ref{fig:figure1}d) shows the calculated force along $x$ in $(x,v_x)$ space for the $^{112}$Cd isotope and magnet set I. The detuning of the trapping laser frequency from resonance is $\Delta_{\mathrm{MOT}}/(2\pi) = -150$ MHz, the detuning of the slower laser frequency is $\Delta_{\textrm{slower}}/(2\pi) = -990$ MHz and its polarization is chosen as the correct circular handedness for Zeeman slowing along $x$. The slowing laser produces a region of force which is always negative (slowing), but whose resonant velocity follows the Zeeman shift of the $F'=1,m_{F'} = +1$ Zeeman sub-level of the excited state. The MOT laser light generates the region of force within a few mm of the trap center. To simulate capture from the atomic beam, we overlay trajectory simulations for atoms launched $20$~cm from the MOT center, with a range of initial velocities $v_x$ typical for a cryogenic buffer gas beam. The white trajectories are successfully captured in the MOT. The simulations predict that the capture velocity of the MOT is $246$~m/s with and $105$~m/s without the Zeeman slower, respectively.

\section{Results}

\subsection{Trap loading and MOT properties}

\begin{figure*}
    \centering
        \includegraphics[width = \textwidth]{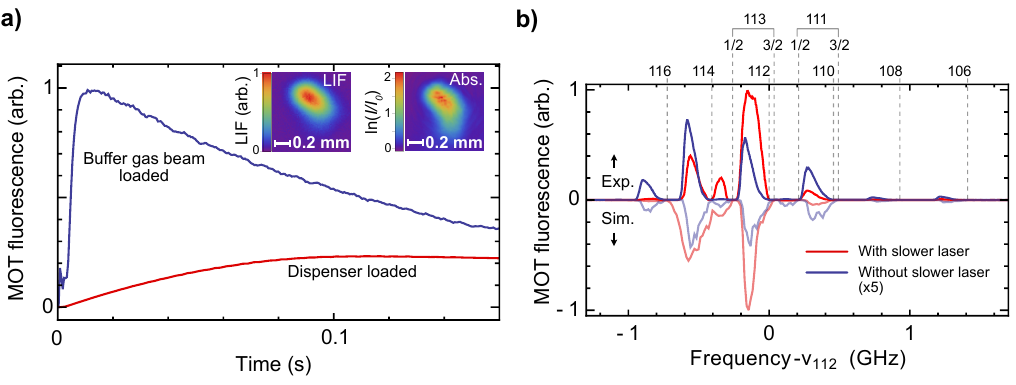}
    \caption{a) Blue: Loading the MOT with $^{112}$Cd atoms from a cryogenic helium buffer gas beam using a Cd ablation target of natural isotopic abundance. Red: Loading a MOT from a dispenser containing Cd with natural abundance. The trap light intensity and detuning is the same in the two cases. The insets show fluorescence (`LIF') and absorption (`Abs.') images of the buffer gas beam loaded MOT. Using the absorption image, we determine the total number of atoms in the MOT to be $1.1(2)\times 10^7$ using a total trap light intensity of 2.2 W/cm$^2$. b) Spectrum showing fluorescence from the MOT with and without a slowing laser and $\Delta_{\mathrm{slower}}=-700$~MHz with respect to the $^{112}$Cd transition. Experimental data are shown pointing upwards, and the results of Monte Carlo simulations are shown pointing downwards. Dashed lines mark isotope resonance frequencies; the $F'$ quantum numbers are shown for the fermionic isotopes.}
    \label{fig:figure2}
\end{figure*}

Figure \ref{fig:figure2}a) shows fluorescence traces recorded with the PMT during MOT loading. The blue trace corresponds to loading from the buffer gas beam, with both slowing and trapping light present and beam polarizations optimized for MOT operation. The signal decays with a $1/e $ lifetime of 150 ms, primarily limited by photoionization losses (see Section \ref{sec:photo}). The red trace shows MOT loading from a Cd dispenser and without the use of a slowing laser. In both cases, the trapping laser intensity and detuning $ \Delta_{\mathrm{MOT}} $ are identical. After 10~ms, the number of atoms loaded into the MOT from the buffer gas beam is ten times larger than loaded from the dispenser.


The inset to figure \ref{fig:figure2}a) shows a false colour camera image of $^{112}$Cd MOT fluorescence, integrated over 0 ms\,$<t<$\,50 ms. The parameters used to load the MOT shown in the image are $\Delta_{\mathrm{MOT}}/(2\pi) = -145$~MHz and $s=I/I_{\mathrm{sat}}\approx 2$, where $I_{\mathrm{sat}} =\pi h c \Gamma/3\lambda^3 = 1.08$~W/cm$^{2}$ is the two-level saturation intensity. The number of photons detected by the camera is given by $n_{\mathrm{ph}} = N_{\mathrm{MOT}}t_{\mathrm{exp}} R_{\mathrm{sc}} \alpha$, where $t_{\mathrm{exp}} = 50$~ms is the exposure time, $\alpha = 1.3 \times 10^{-6}$ is the total detection efficiency of the imaging system and camera and $R_{\mathrm{sc}}$ is the mean scattering rate averaged over the trap. For the image in figure \ref{fig:figure2}a), the value of $\alpha$ includes a neutral density filter of optical depth 2.0 to prevent camera saturation. We estimate $R_{\mathrm{sc}}$ using rate equations for atoms in the MOT and assume isotropic emission of the fluorescence. For the bosonic isotopes, there is a single ground state and three excited states, and we find,
\begin{equation}
\begin{aligned}
    R_{\mathrm{sc}} &= \frac{\Gamma_{\mathrm{eff}}}{2}\frac{s_{\mathrm{eff}}}{1+s_{\mathrm{eff}} + 4\Delta_{\mathrm{MOT}}^2/\Gamma^2} \hspace{0.1cm},
\end{aligned}
\end{equation}

\noindent where we have used the effective spontaneous decay rate $\Gamma_{\mathrm{eff}}= 3\Gamma/2$ and the effective saturation parameter $s_{\mathrm{eff}} = 2s/3$ to give an expression analogous to that of a two-level system. Here, we assume that the total scattering rate does not change across the MOT and the total peak intensity is $I=sI_{\mathrm{sat}}$. For the trap parameters for the MOT shown in figure \ref{fig:figure2}a), we find $R_{\mathrm{sc}} = 0.082\Gamma$, about $5\%$ larger than when predicted assuming a two-level system driven by the total intensity. The result is that the camera detects 3.4 photons per atom over the exposure time, and we estimate $N_\textrm{MOT} = 1.2 \times 10^6$.

An alternative method of measuring the number of trapped atoms is via their absorption of a low-intensity probe beam. A false color image of the absorption shadow from a $^{112}$Cd MOT is shown as an additional inset to figure \ref{fig:figure2}a), where the probe beam intensity is $0.1 I_{\mathrm{sat}}$. We image the atoms with the trapping light switched on to prevent rapid expansion of the cloud resulting from the high Doppler temperature (see below) or excessive displacement of the cloud by the probe beam. The probe beam is switched on for 0.2~ms using an acousto-optic modulator. A short exposure time minimizes fluorescence from the MOT beams collected onto the camera. The absorption at pixel $i$ on the camera, $\mathcal{A}_i$, is calculated as

\begin{equation}
    \mathcal{A}_i = -\mathrm{ln}(I_i/I_{i,0}) = n \sigma_{\mathrm{abs}}
\end{equation}

\noindent where $I_i$ and $I_{i,0}$ are the pixel intensity values with and without atoms in the MOT, $n$ is the number of atoms per unit area and $\sigma_{\mathrm{abs}}$ the absorption cross-section. The resonant absorption cross-section $\sigma_{0} = 3\lambda^2/2\pi = 2.5\times 10^{-10}$~cm$^2$ for the Cd bosons, around one order of magnitude smaller than for the D2 transitions in alkali atoms. Broadening due to the velocity distribution and the magnetic fields sampled in the trap is neglected (i.e. $\sigma_{\mathrm{abs}}= \sigma_0$), setting a lower bound to $N_{\mathrm{MOT}}$. To arrive at our estimate of the number of atoms, we calculate the sum $p^2/\sigma_{\mathrm{abs}}\Sigma_i \mathcal{A}_i$ over the absorption image, where $p=13~\mu$m is the pixel size, which yields $N_{\mathrm{MOT}} = 1.1(2)\times10^7$ (with an uncertainty of $10\%$ on the size of the MOT). This number is about a factor of nine larger than the value obtained from fluorescence, and we consider the absorption measurement superior; it does not rely on the precise knowledge of the laser intensity, efficiency of the collection optics and is not affected by radiation trapping effects. In addition, the total intensity seen by the atoms can be significantly lower than the measured peak intensity of the MOT beams. Due to the relatively small size of the MOT beams and high magnetic field gradient, efficiently overlapping them and aligning them to the magnetic field zero is challenging and the exact location where the MOT forms is not known.

Figure \ref{fig:figure2}b), upper panel, shows the fluorescence signal from the MOT between 100~ms$\,<t<\,$200~ms versus the trapping laser frequency. 
The solid red (blue) lines are taken with (without) the slower laser applied, with $\Delta_{\textrm{slower}}/(2\pi)=-700$~MHz which is optimal for trapping $^{112}$Cd. The dashed lines indicate transition frequencies of the different Cd isotopes, and we indicate the excited state total angular momentum quantum number $F'$ for the fermions. All bosonic isotopes are trapped over a detuning range $-2\Gamma<\Delta_{\mathrm{MOT}}<0$ from their respective transition resonance. Pointing downwards in figure \ref{fig:figure2}b) are the results of a Monte Carlo trajectory simulation of the MOT loading. The initial forward velocity $v_x$ in the simulations is sampled from a Gaussian velocity distribution representative of the atomic beam (the most probable velocity, $\bar{v}_x=130$\,m/s, with a full width at half maximum of $60$\,m/s), and the transverse velocity (perpendicular to $x$), $\left(v_y^2+ v_z^2\right)^{1/2}$ is sampled from a top-hat distribution centered around 0 m/s with a range of 14 m/s. To account for experimental misalignment, the slower beam within the simulation is angled by 1.5 mrad relative to the molecular beam axis. Each simulated data point is the result of $5\times 10^4$ trajectories per Cd isotope, and we count the number of atoms that satisfy the conditions $x^2 + y^2 + z^2 < 2\times{}10^{-2}$ mm$^2$, $v_x^2 + v_y^2 + v_z^2 < 0.2$ m$^2$/s$^2$. The simulations qualitatively reproduce the experimental observations, predicting that the slowing laser enhances the number of $^{112}$Cd atoms in the MOT about tenfold over the range of detunings used in the experiment. Since the $^{110}$Cd ($^{112}$Cd) resonance almost exactly coincides with the $^{111}$Cd ($^{113}$Cd) resonance with $F'=3/2$, we rely on the simulations to estimate the ratio of the bosonic to fermionic isotopes at the optimum trap laser detuning. The simulations predict that with $\Delta_{\mathrm{MOT}}$ and $\Delta_{\mathrm{slower}}$ optimised for $^{112}$Cd, there is a negligible fraction of $^{113}$Cd trapped, but if $\Delta_{\mathrm{MOT}}/(2\pi)=-350$~MHz with respect to the $^{112}$Cd resonance, $^{113}$Cd can be loaded into the trap. The slower laser significantly increases the number of $^{113}$Cd atoms loaded at this value of $\Delta_{\mathrm{MOT}}$, since our choice of $\Delta_{\mathrm{slower}}$ is suitable for slowing both $^{112}$Cd and $^{113}$Cd. In contrast, $\Delta_{\mathrm{slower}}$ is too large to significantly improve the loading of $^{111}$Cd and actively inhibits loading $^{116}$Cd into the trap.    


\subsection{Trap frequency and damping constant}
\begin{figure}
    \centering
   
    \includegraphics[]{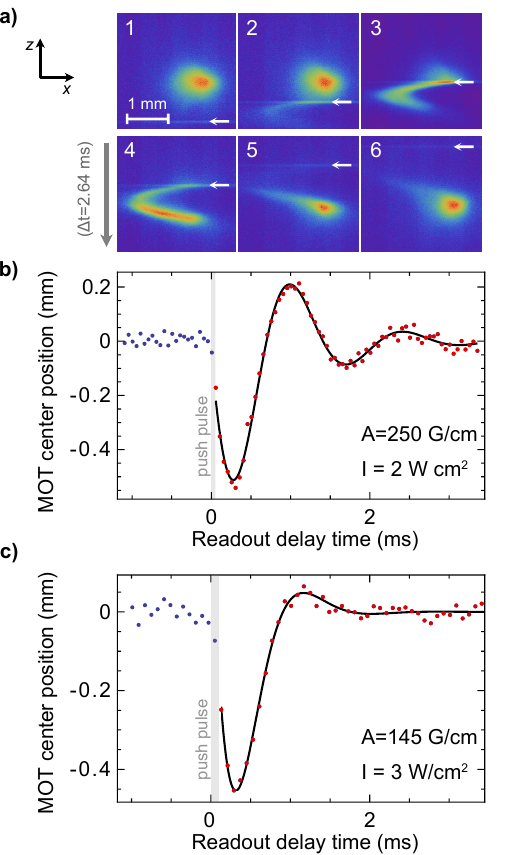}
    \caption{Center-of-mass oscillations of the atomic cloud following a momentum kick along the $-x$ direction applied via a short push laser pulse. a) Fluorescence images showing the oscillatory motion for a large excursion (using magnet set II). The delay between the push pulse and the start of camera readout is incremented in steps of 0.42 ms across the sequence. Due to the rolling shutter, each image exhibits a vertical time gradient: the 10 $\mu$s readout time per row causes the top of each image to be captured earlier than the bottom. The push pulse appears as a narrow horizontal band (highlighted with a white arrow) and moves upwards through the image sequence as the delay increases. b) and c) Center-of-mass position of the cloud, extracted from a narrow horizontal slice of each image, plotted versus the camera delay time. Two different MOT configurations are shown. Blue (red) points correspond to measurements taken before (after) the application of the push pulse. The solid black lines are fits to a damped harmonic oscillator model described in the main text.}
    \label{fig:pushBeam}
\end{figure}
Similarly to MOTs of Ca \cite{Grünert2002}, Mg \cite{Loo2004}, and Sr \cite{Xu2002}, the simple structure of the Cd bosons means that motion in the trap should be well described by Doppler cooling theory. This contrasts the MOTs of alkali atoms where sub-Doppler cooling forces lead to a substantial increase in the damping constant and complicate the modeling of in-trap dynamics \cite{Kohns1993}. To measure the trap frequency and damping constant of our MOT, we push the atoms to a velocity of $3-5$~m/s along $x$ using a 50~$\mu$s pulse of near-resonant light from the slower laser. We then observe damped harmonic motion of the atomic cloud via its fluorescence image on the camera. Our temporal resolution is set by the minimum readout time for each row of camera pixels of 10~$\mu$s. The readout of the region covering the atomic cloud takes 500~$\mu$s in total, approximately half the oscillation period in the trap. As a result, each row of pixels captures part of the cloud with a specific time delay relative to the push laser pulse, and the motion is visible within a single frame. Figure \ref{fig:pushBeam}a) shows a set of camera images that illustrate this effect, where the delay between the push laser pulse and the camera readout is increased in $0.42$~ms steps. The push pulse appears as a horizontal stripe in each image, marked by the white arrow. Image one uses a readout delay such that the push pulse appears at the bottom (i.e. late in the exposure), meaning that the atoms appear undisturbed. In images two to six, the push laser pulse occurs during or before the atomic fluorescence is recorded, leading to a bright horizontal stripe of fluorescence in the image, with the subsequent motion of the cloud being captured. 

To extract the motion of the cloud from the camera images, we select three pixel columns near the cloud center and vary the readout time of this region relative to the application of the push laser beam. Figure \ref{fig:pushBeam}b) shows the center-of-mass within the range of interest as a function of time, using magnet set I. Damped oscillations following the push laser pulse are clearly visible. The black solid lines show fits to the equation $x(t) = a \exp[-\beta t/2]\cos\left[\sqrt{\omega_{\mathrm{trap}}^2 -\beta^2/4} t + \phi_0\right]$, where $a$ is the amplitude of the oscillation and $\phi_0$ is a phase. From the fit we extract the trap angular frequency $\omega_{\mathrm{trap}} = 4.5(1)\times10^3$ s$^{-1}$ and the damping constant $\beta=2.5(1)\times10^3$ s$^{-1}$. We can calculate these constants using a rate equation model that includes the effect of the vertical (non-restoring) beams using the following expressions: 



\begin{equation}
    \omega_{\mathrm{trap}}^2=-\frac{4}{3m}\frac{\Delta_{\mathrm{MOT}}}{\Gamma}\frac{\mu_B A k s}{(\xi+2s/3)(\xi+s/6)},
    \label{eqn:trapFreq}
\end{equation}

\begin{equation}
    \beta = -\frac{4}{3m}\frac{\Delta_{\mathrm{MOT}}}{\Gamma}\frac{\hbar k^2 s(\xi+s/12)}{\xi(\xi+2s/3)(\xi+s/6)}.
    \label{eqn:trapDamping}
\end{equation}

\noindent Here, $A$ is the magnetic field gradient along the axis of motion ($x$), $k = 2\pi/\lambda$ is the angular wavenumber of the trapping light, $m$ is the $^{112}$Cd atomic mass and we have defined $\xi\equiv1+4\Delta_\mathrm{MOT}^2/\Gamma^2$ for brevity. Within the rate equation model, $\omega_{\mathrm{trap}}$ and $\beta$ are independent of the direction of the excursion in the $x$-$y$ plane. 

For our trap parameters, equations \eqref{eqn:trapFreq} and \eqref{eqn:trapDamping} predict a trap frequency of $ \omega_{\mathrm{trap}} = 11.2 \times 10^3 \,\mathrm{s}^{-1} $ and a damping constant of $ \beta = 16.0 \times 10^3 \,\mathrm{s}^{-1} $. Both values significantly exceed the experimentally observed ones. This discrepancy is most likely due to imperfect spatial overlap of the trapping beams, non-ideal beam balancing, polarization imperfections, and the fact that atoms undergo excursions from the trap center, where they experience a reduced average intensity. These are common challenges in MOTs operating in the DUV. The lower measured values of $ \omega_{\mathrm{trap}} $ and $ \beta $ indicate a reduced scattering rate, consistent with the underestimation of atom number in the fluorescence measurement discussed above. Notably, for our parameters, $ \beta $ is particularly sensitive to the exact value of $ \Delta_{\mathrm{MOT}} $.

An important feature of equations \eqref{eqn:trapFreq} and \eqref{eqn:trapDamping} is that the ratio $ \omega_{\mathrm{trap}}^2/\beta = A\mu_B\xi/(k\hbar(\xi + s/12)) $ depends only weakly on both the MOT beam intensity and detuning. Using a magnetic field gradient of $ A = 250 \,\mathrm{G/cm} $, we obtain a theoretical value of $ \omega_{\mathrm{trap}}^2/\beta = 7.9 \times 10^3 \,\mathrm{s}^{-1} $, which is in good agreement with the experimentally measured value of $ 8.1(5) \times 10^3 \,\mathrm{s}^{-1} $.
\begin{table}
\caption{\label{tab:trap_properties} Trap angular frequency $\omega_{\textrm{trap}}$ and damping constant $\beta$ are determined from the fits shown in Figure \ref{fig:pushBeam}b) and \ref{fig:pushBeam}c) for two different sets of permanent magnets. Theoretical values derived from  Doppler theory, using equations \ref{eqn:trapFreq} and \ref{eqn:trapDamping} are given. The discrepancy between theory and experiment is discussed in the text.}
\begin{ruledtabular}
    \begin{tabular}{lllll}
        I (W/cm$^2$) & A (G/cm) & $\omega_{\textrm{trap}}$ (10$^3$s$^{-1}$) & $\beta$ (10$^3$s$^{-1}$) \\
        \hline
        2 & 250 & $4.5(1)$ & $2.5(1)$\\
        Theory & & $11.2$ & $16$\\
        3 & 145  & $4.5(1)$ & $5.3(1)$\\
        Theory & & $10.1$ & $22.5$\\

    \end{tabular}
    \end{ruledtabular}
\end{table}
We repeated the trap oscillation measurements with the second set of magnets for which $A = 145$~G/cm, shown in Figure \ref{fig:pushBeam}c), using the same value of $\Delta_{\mathrm{MOT}}$ and a slightly higher total peak intensity $I = 3$~W/cm$^2$, finding $\omega_{\mathrm{trap}} = 4.5(1)\times10^3$s$^{-1}$ and $\beta=5.3(1)\times10^3$s$^{-1}$. Comparing the data in Figure \ref{fig:pushBeam}b) and c), the trap frequencies are equal, while the damping constant is a factor of two larger. From equations \ref{eqn:trapFreq} and \ref{eqn:trapDamping}, we expect a slight drop in the trap frequency ($\sim 10\%$) and a damping constant that is larger by $40\%$.  Table \ref{tab:trap_properties} summarizes the results.

From the trap frequency measurements, we estimate the temperature of the atoms in the MOT. Using the equipartition theorem we relate $T=m\omega_{\mathrm{trap}}^2 \langle x^2 \rangle /k_B$, where $\langle x^2 \rangle $ is the mean square width of the cloud along the axis in which $\omega_{\mathrm{trap}}$ was measured. From the images in Figure \ref{fig:figure2}b), we estimate $T=6.3$~mK, and a peak density at the trap center of $\rho = 2.5 \times 10^{11}$cm$^{-3}$, with a peak phase-space density of $2\times10^{-9}$.
These conditions are well suited for transfer into a second-stage MOT operating on the 66.6 kHz wide $^1S_0\rightarrow\,^3P_1$ intercombination transition at 326 nm \cite{Katori2019, gibble_laser-cooling_2024}. When our loading method is combined with current-carrying coils for optimum transfer 
and cooling efficiency to this second stage cooling, we expect a significant increase in the phase-space density compared to the current state-of the art.

%


%
\subsection{\label{sec:photo}Loss mechanisms}
\begin{figure}
    \centering
        \includegraphics[width=\columnwidth]{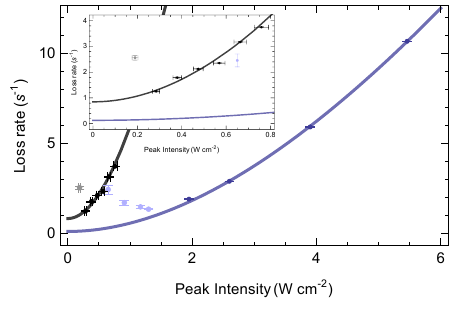}
    \caption{The loss rate from the MOT is dominated by two-photon ionization. Data from two independent experiments are shown: At FHI (blue) we determine the lifetime of a beam-loaded MOT ($\Delta_{\textrm{MOT}}/(2\pi)=-150$ MHz) and at ICL (black) we measured loading rates of a dipsenser-loaded MOT ($\Delta_{\textrm{MOT}}/(2\pi)=-100$ MHz) as a function of the laser intensity. The light blue and gray points are excluded from the fit as the MOT became unstable. The solid curves represent fits to the data using equation \ref{eqn:ionloss}. The cross-sections determined from this data are summarized in table \ref{tab:ionisation}. The observed difference in the two datasets can be partially attributed to the different detunings; however, the extracted cross sections still differ by approximately three standard deviations, underscoring the challenges of determining the light intensity in the MOT.}
    \label{fig:lossrate}
\end{figure}
Similar to Mg atoms the intersystem radiative decay channel $^1P_1\rightarrow\,^3P_j$ is highly unlikely and has not been measured or detected so far. The dominant loss mechanism in our MOT arises from resonant two-photon ionisation via the $^1P_1$ state, as the energy of two 229 nm photons exceeds the ionization potential of Cd. The corresponding loss rate is 
\begin{equation}
   A_{\mathrm{ion}} = I \frac{\sigma_{\mathrm{ion}} \rho_{ee}}{\hbar \omega}, 
   \label{eqn:ionloss}
\end{equation}
where  $\sigma_{\mathrm{ion}}$ is the photoionization cross section of the $^1P_1$ state and $\rho_{ee}$ is the spatially averaged excited-state fraction. The total atom loss rate is modeled as $A = A_{\mathrm{ion}} + A_0$, with $ A_0$ representing loss from background gas collisions, assumed to be independent of the MOT parameters \cite{madsen_measurement_2002}.
Figure \ref{fig:lossrate} shows the measured total loss rate as a function of the laser intensity (points) and corresponding fits (solid curves) using equation \ref{eqn:ionloss}. The fit allows us to determine the absolute ionization cross-section $\sigma_{\mathrm{ion}}$. We measured the cross-section in two independent experiments, using distinct MOT parameters, laser-intensity profiles and atom sources. The first experiment (shown in blue), conducted at FHI, used a beam-loaded MOT, and a detuning of $ \Delta_{\mathrm{MOT}}/(2\pi) = -145 $ MHz with peak intensities $I>1.5\ \mathrm{W/cm}^2$.
\begin{table}[h]
\caption{\label{tab:ionisation}
Measured photoionization cross section of the $ ^1P_1$ state of Cd.
The first two values are extracted from fits to the data shown in Figure \ref{fig:lossrate}, and are compared to the previously reported measurement from \cite{Brickman2007}. Quoted uncertainties represent combined statistical and systematic errors (in the final digit). The laser intensity range used for each measurement is also indicated.}
\begin{ruledtabular}
    \begin{tabular}{lll}
        Method & $\sigma_\textrm{ion}$ $(10^{-16} \,\textrm{cm}^2)$ & Intensity (W/cm$^2$) \\
        \hline
        Beam-loaded & 0.2(2) & $2-6$ \\
        Dispenser-loaded & 0.8(2) & $0.1-0.8$ \\
        Dispenser-loaded \cite{Brickman2007} & 2(1) & $0.035-0.07$ \\
    \end{tabular}
    \end{ruledtabular}
\end{table}
The second measurement, performed at ICL, used a dispenser-loaded MOT (shown in black) with a detuning of $ \Delta_{\mathrm{MOT}}/(2\pi) = -100\,\mathrm{MHz}$ and a laser intensity range of $0.2 < I < 0.8\,\mathrm{W/cm}^2$. Here, we determine the loading rate of the MOT and fit the model $N(t)=N_{ss}(1-e^{-At})$, where $N_{ss}$ is the steady state MOT population. Table \ref{tab:ionisation} summarizes the results and compares our measurements to those reported in reference \cite{Brickman2007}. While the statistical uncertainties from the fits are small, the total uncertainty in our measurement is dominated by systematic effects related to the MOT geometry, laser beam alignment, and the determination of the intensity distribution across the atomic cloud. To address this, we implemented camera-based beam profiling before and after the measurement to accurately determine the spatial intensity profile. This allowed us to relate the peak intensity to the total beam power and apply corrections for deviations from an ideal Gaussian mode. Despite this, evidence from trap frequency measurements and MOT fluorescence indicates that the atoms experience a lower average intensity than inferred from beam profile and power measurements. This is likely due to imperfect overlap between the MOT and the high-intensity region of the beams, which further contributes to the uncertainty in estimating the effective intensity. Additionally, the power meter carries a stated uncertainty of 10\%. Overall, our results are consistent with the earlier determination by Brickman et al. \cite{Brickman2007}. Taken together, the data reflect systematic variability across experiments.
\begin{figure*}
    \centering
    \includegraphics[width=2\columnwidth]{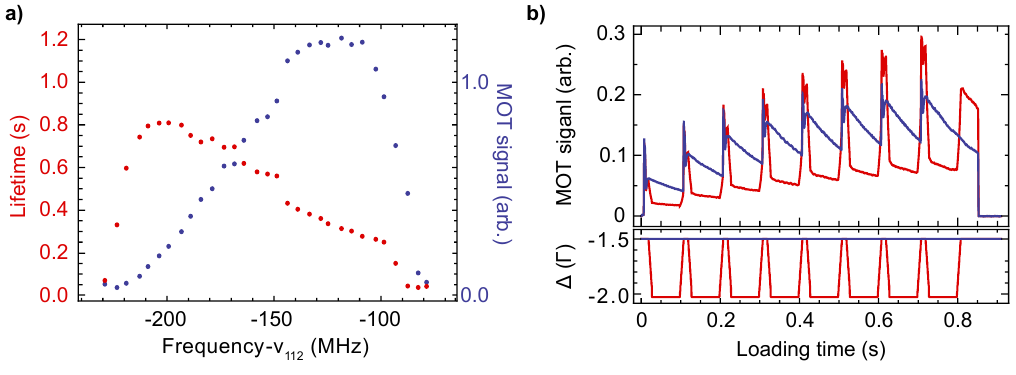}
    \caption{Control of photoionization losses from the MOT. a) $^{112}$Cd MOT lifetime (red points, left axis) and MOT fluorescence (blue points, right axis) versus the trap light detuning $\Delta_{\mathrm{MOT}}$. b) Accumulation of multiple atomic pulses into the trap, for two different configurations of the MOT. The upper panel shows the MOT fluorescence as eight successive atomic pulses are loaded into the MOT at 100~ms intervals. The lower panel shows the time dependence of $\Delta_{\mathrm{MOT}}$ for the two experiments. Detuning the trap light further from resonance after loading (red) reduces photoionization losses, increasing the number of trapped atoms compared to fixing the detuning near the optimum for loading (blue).}
    \label{fig:lifetime}
\end{figure*}

To reduce photoionization losses in the MOT, we explored strategies that minimize the excited-state fraction $\rho_{ee}$ after loading. One approach is to reduce the trap laser intensity by lowering the RF drive power to an acousto-optic modulator. However, this leads to unwanted shifts in the beam position at the MOT location due to small changes in the diffraction angle, rendering the method ineffective. As an alternative, we implemented a switched-detuning scheme in which the trapping laser frequency is rapidly detuned further from resonance immediately after loading, while keeping the intensity $I$ constant. This reduces $\rho_{ee}$ without affecting beam alignment. Figure \ref{fig:lossrate}a) shows the MOT number and lifetime for $^{112}$Cd, as a function of the trap laser detuning $\Delta_{\mathrm{MOT}}$. The initial atom number is optimized at $\Delta_{\mathrm{MOT}}/(2\pi)=-130$~MHz, whereas the longest lifetime is achieved at $-210$~MHz. Based on this, we load the MOT at $\Delta_1 = -1.5\Gamma$ and then linearly ramp to $\Delta_2 = -2\Gamma$ using a voltage-controlled adjustment of the active mirror in the Ti:Sa laser.

The benefit of this scheme becomes clear when accumulating atoms from multiple pulses, as shown in Figure \ref{fig:lifetime}b). Here, we compare two approaches: fixed detuning (blue) versus the switched-detuning scheme (red), for a sequence of eight atomic pulses spaced by 100 ms. The lower panel shows the time-dependent trap light detuning $\Delta_{\mathrm{MOT}}$ for each loading scheme. Shortly after loading each atomic pulse, the two schemes have identical values of $\Delta_{\mathrm{MOT}}$, and the trap fluorescence clearly illustrates the benefit of the switched-detuning scheme. At $t=0.85$~s, following the loading sequence, the trap conditions are again identical. We find that the switched-detuning scheme delivers a factor 1.6 more atoms to the MOT compared with fixed detuning, and a factor 3.5 more atoms than loading with a single pulse. Importantly, loading the trap with short pulses enables this gain in efficiency, and the method is ineffective in a dispenser loaded MOT where loading is continuous. The scheme could be further enhanced by increasing the repetition rate of the atomic beam source, by using a high-repetition-rate YAG laser in combination with a closed liquid helium reservoir to manage heating during operation \cite{PhysRevA.102.041302}. Such an upgrade would enable even higher MOT loading rates and is expected to result in a substantial increase in the final atom number.


\section{Conclusion \& Discussion}
We have demonstrated a high-density, deep-ultraviolet magneto-optical trap of cadmium atoms, loaded from a cryogenic helium buffer-gas beam. The MOT captures up to $ 1.1 \times 10^7$ atoms in 10 ms from a single pulse at a loading rate exceeding $ 10^9\,\mathrm{s}^{-1}$, resulting in a peak density of $2.5 \times 10^{11}\,\mathrm{cm}^{-3}$. This performance is enabled by the strong radiation pressure force at 229 nm and the use of a short, integrated Zeeman slower. We have characterized photoionization losses and demonstrated a switched-detuning strategy that increases the MOT atom number by a factor of 3.5. Further improvements in the MOT number can be expected by using an isotope-enriched target and a higher source repetition rate. Despite significant loss channels due to two photon ionization by the trapping light, the performance of our Cd MOT in terms of atom number and density approaches that of state-of-the-art MOTs of Mg \cite{madsen_generation_2002, riedmann_beating_2012}, Ca \cite{curtis_quenched_2001,binnewies_doppler_2001,grunert_sub-doppler_2002}, Sr \cite{katori_magneto-optical_1999,courtillot_efficient_2003,nosske_two-dimensional_2017}, Yb \cite{honda_magneto-optical_1999,letellier_loading_2023}, and Hg \cite{hachisu_trapping_2008,petersen_doppler-free_2008,Lavigne2022}.

Our approach is readily adaptable to other atomic species that suffer from high loss rates in the main cooling cycle and offers the potential to accelerate MOT loading significantly. In contrast to conventional oven-based setups, ablation-based sources require substantially less material of the target species, making it well suited for experiments with rare isotopes and enriched targets.

The high number and density achieved opens the possibility to study Bose-Einstein condensates of Cd and enables applications ranging from quantum simulation, precision isotope shift measurements, atom interferometry and exploration of novel ion-neutral hybrid systems.

Narrow-line laser cooling and magic-wavelength optical trapping of Cd have recently been demonstrated \cite{Katori2019, gibble_laser-cooling_2024}, establishing the core ingredients required for operating a cadmium optical lattice clock. These results, combined with the high atom number and fast loading rate demonstrated here, position Cd as a promising candidate for future clock development.

The high brightness and short pulse duration of the buffer-gas beam make it ideally suited for loading molecular species with limited photon scattering budgets. In particular, the stable AlF molecule, like Cd, features a DUV cooling transition and experiences comparable radiation pressure forces. Notably, the brightness of our $^{112}$Cd beam is comparable to that of the AlF beam we reported recently \cite{hofsass2021optical, wright_cryogenic_2023}, suggesting that similar MOT loading strategies can be applied. We have successfully benchmarked the DUV laser cooling system using Cd and are now ready to extend this approach to AlF. Realizing a MOT of AlF will require two additional DUV laser systems to re-pump the population from higher vibrational states in the ground electronic state. The modular design of the setup presented here enables rapid switching between Cd and AlF operation, allowing parallel development of atomic and molecular cooling techniques within the same apparatus.

\begin{acknowledgments}
We gratefully acknowledge the expert technical support provided by the mechanical and electronic workshops of the Fritz Haber Institute. This work was supported by the European Research Council (ERC) under the European Union’s Horizon 2020 Research and Innovation Programme (Grant Agreement No. 949119, project ``CoMoFun''), and by the European Commission through project ``UVQuanT'' (Grant Agreement No. 101080164).
\end{acknowledgments}

\bibliography{references}

\end{document}